\documentclass[10pt]{article}
\usepackage[a4paper, left=0.8in, right=0.8in, top=0.8in, bottom=1in]{geometry}
\usepackage{multicol}
\usepackage{amsmath,amssymb}
\usepackage{graphicx}
\usepackage{float}
\usepackage{cite}
\usepackage{multirow}
\usepackage{wrapfig}
\usepackage{caption}
\usepackage{siunitx}
\usepackage{hyperref}

\allowdisplaybreaks

\title{An Open Source Python Package to Simulate\\Micro Thermoelectric Generators}
\author{D. Beretta\thanks{Corresponding Author: davide.beretta@uzh.ch} \\
        University of Zurich \\}
        
\date{\small December 8, 2024}

\begin{document}

\maketitle

\begin{abstract}
This article presents an open-source Python package for simulating micro-thermoelectric generators, based on the work by D. Beretta et al. (Sustainable Energy Fuels, 2017). Featuring a user-friendly graphical user interface and robust computational capabilities, the tool is designed for use by scientists, researchers, and engineers to analyze and optimize device designs. The software calculates key performance metrics such as power, efficiency, electrical resistance, open circuit voltage, and short circuit current per unit of device area, based on the device design and material properties. The full source code is available for download on GitHub, enabling further customization.
\end{abstract}

\begin{table*}[hb]
    \small
    \centering
    \begin{tabular}{ll}
    \hline
    \textbf{Symbol} & \textbf{Description} \\
    \hline
    $\alpha_p, \alpha_n$ & Seebeck coefficient of p- and n-type materials (\SI{}{\volt\per\kelvin}) \\
    $\alpha_{pn}$ & Thermocouple Seebeck coefficient (\SI{}{\volt\per\kelvin}) \\
    $\sigma_p, \sigma_n$ & Electrical conductivity of p- and n-type materials (\SI{}{\siemens\per\meter}) \\
    $\rho_p, \rho_n$ & Electrical resistivity of p- and n-type materials (\SI{}{\ohm\times\meter}) \\
    $\rho_{pn}$ & Average electrical resistivity of a thermocouple (\SI{}{\ohm\times\meter}) \\
    $\kappa_p, \kappa_n, \kappa_i$ & Thermal conductivity of p-type, n-type, and insulating material (\SI{}{\watt\per\meter\per\kelvin}) \\
    $\pi_p, \pi_n$ & Peltier coefficient of p- and n-type materials (\SI{}{V}) \\
    $\pi_{pn}$ & Thermocouple Peltier coefficient (\SI{}{V}) \\
    $zT$ & Thermoelectric figure of merit (dimensionless) \\
    $\dot{q}_h, \dot{q}_c$ & Hot- and cold-side heat flux (\SI{}{\watt\per\meter\squared}) \\
    $\dot{q}_{\pi,h}, \dot{q}_{\pi,c}$ & Peltier heat flux at the hot and cold sides (\SI{}{\watt\per\meter\squared}) \\
    $\dot{q}_j$ & Joule heat flux (\SI{}{\watt\per\meter\squared}) \\
    $h_{r,h}, h_{r,c}$ & Heat transfer coefficient of the hot and cold reservoir (\SI{}{\watt\per\meter^2\per\kelvin}) \\
    $h_{s,h}, h_{s,c}$ & Heat transfer coefficient of the hot and cold substrate (\SI{}{\watt\per\meter^2\per\kelvin}) \\
    $h_{pni}$ & Heat transfer coefficient of the thermocouple unit (\SI{}{\watt\per\meter^2\per\kelvin}) \\
    $T_{r,h}, T_{r,c}$ & Temperature of the hot and cold reservoir (\SI{}{\kelvin}) \\
    $T_{s,h}, T_{s,c}$ & Temperature of the hot and cold substrate (\SI{}{\kelvin}) \\
    $T_h, T_c$ & Thermocouple hot- and cold-side temperature (\SI{}{\kelvin}) \\
    $n$ & Number density of thermocouples (\SI{}{\per\meter\squared}) \\
    $A_p, A_n, A_i$ & Cross-sectional area of p-type, n-type, and insulating material (\SI{}{\meter\squared}) \\
    $l$ & Thermocouple length (\SI{}{m}) \\
    $R_p, R_n$ & Electrical resistance of a p-type and n-type leg (\SI{}{\ohm}) \\
    $R_{pn}$ & Thermocouple electrical resistance (\SI{}{\ohm})\\
    $R$ & Device internal electrical resistance (\SI{}{\ohm}) \\
    $R_{load}$ & Load resistance (\SI{}{\ohm}) \\
    $f$ & Fill factor (dimensionless) \\
    $p$ & Output power per unit of device area (\SI{}{\watt\per\meter\squared}) \\
    $\eta$ & Conversion efficiency (dimensionless) \\
    $V_{oc}$ & Device open-circuit voltage (\SI{}{\volt)} \\
    $v_{oc}$ & Open-circuit voltage per unit of device area (\SI{}{\volt\per\meter\squared}) \\
    $I_{sc}$ & Device short-circuit current (\SI{}{\ampere}) \\
    $i_{sc}$ & Short-circuit current per unit of device area (\SI{}{\ampere\per\meter\squared}) \\
    $I_{cc}$ & Device closed-circuit current (\SI{}{\ampere}) \\
    $i_{cc}$ & Closed-circuit current per unit of device area (\SI{}{\ampere\per\meter\squared}) \\
    $r$ & $\mu$TEG electrical resistance per unit of device area (\SI{}{\ohm\per\meter\squared}) \\
    \hline
    \end{tabular}
    \label{tab:symbols}
\end{table*}

\begin{multicols}{2}
\section{Introduction}
Micro Thermoelectric Generators ($\mu$TEGs) are solid-state devices designed to convert heat into electrical energy through the Seebeck effect.\cite{Beretta.Caironi.Materials} Traditional $\mu$TEGs consist of several thermocouples fabricated in electrical series and thermal parallel, between two electrically insulating substrates, such that a temperature difference between the substrates drives an electric current. Typically, $\mu$TEG are designed to operate under small temperature differences, around room temperature, where modest heat fluxes ensure that the hot reservoir is not depleted. With theoretically infinite reservoirs, $\mu$TEGs are optimized to maximize the generated electrical power, with little regard for conversion efficiency. This optimization requires the thermal resistance of the device to match the thermal resistance of the device's coupling to the reservoirs — a condition known as thermal matching, typically resulting in thermocouples on the order of micrometers in length, hence the name micro-TEGs or $\mu$TEGs.\cite{Beretta.Caironi.Sustainable,Apertet.Lecoeur.Journal} 
$\mu$TEGs are particularly suited for low-power electronics applications, such as battery charging. Despite their potential, the design and optimization of these devices are often hindered by the complexity of simulating their performance, as the governing equations are multi-physics and non-linear, and cannot be solved analytically. This challenge forces researchers and engineers to develop custom code, which can be time-consuming and error-prone, especially for those without extensive programming experience. Alternatively, linearized models are often used, but these tend to overestimate the device’s performance, by more than 10\% in the application ranges of interest \cite{Beretta.Caironi.Sustainable}.
To address this gap, this work presents a user-friendly, open-source Python package based on a lumped-element model to simulate $\mu$TEGs. Building on the work of Beretta et al.,\cite{Beretta.Caironi.Sustainable} the package allows users to calculate key parameters such as power, conversion efficiency, open-circuit voltage, short-circuit current, and electrical resistance per unit of device area as functions of thermocouple length, device geometry, material properties and temperature difference. Featuring a simple and intuitive graphical interface, the package is accessible to both experienced users and those with limited programming experience for performing basic thermoelectric simulations. It is available via pip, can be executed from the command line, and its full source code is hosted on GitHub for transparency and further customization.\cite{beretta_2024_12802052} By simplifying the simulation process, this Python package facilitates rapid prototyping and performance evaluation—an invaluable step before progressing to more advanced Finite Element Method (FEM) multi-physics simulations.

\section{Model}
\begin{figure*}[tb]
    \centering
    \includegraphics[width=0.5\linewidth]{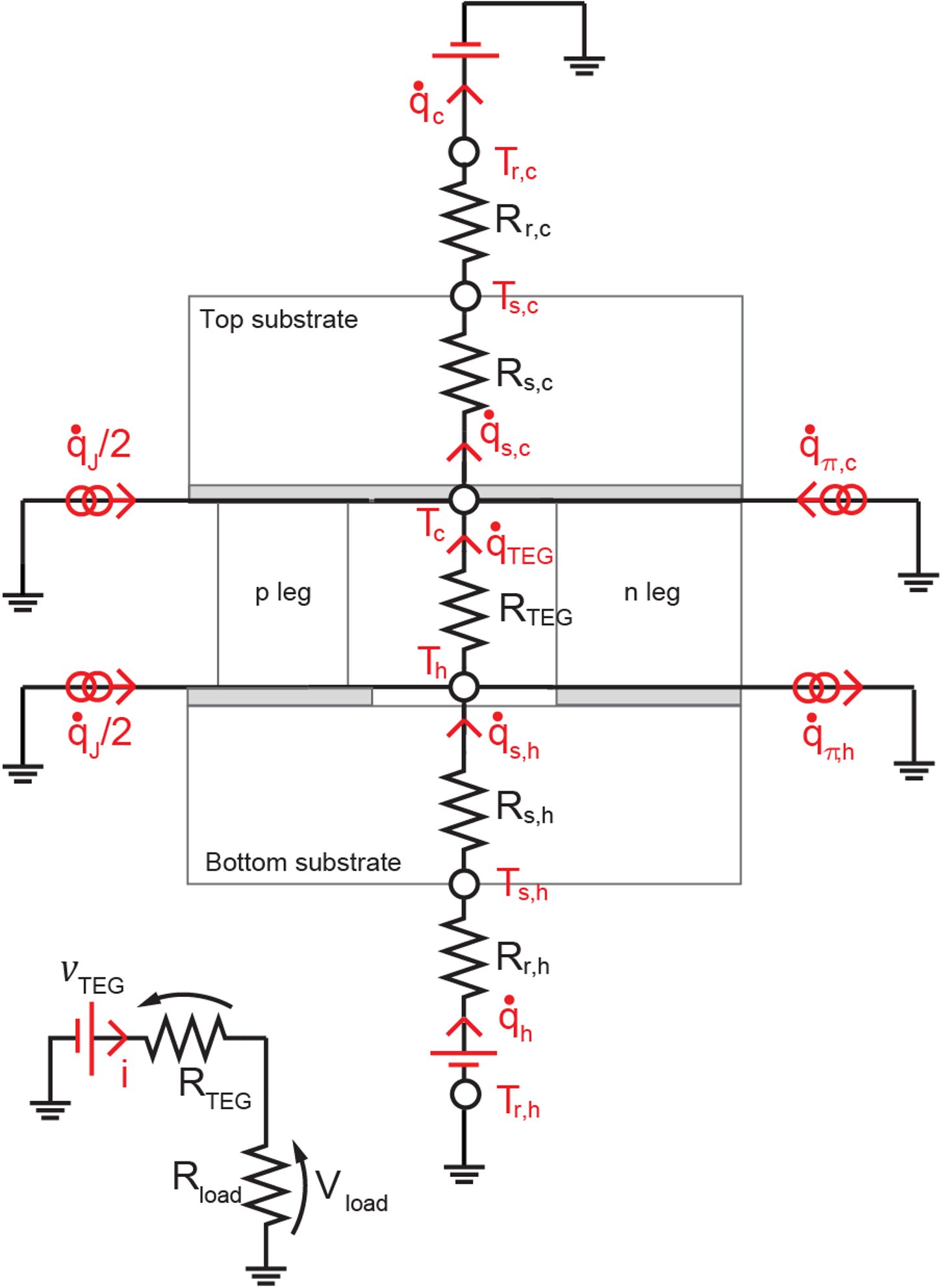}
    \caption{Electrical and thermal schematics of a $\mu$TEG consisting of a single thermocouple. The thermal model excludes the metallic interconnections, assuming their contribution is negligible, as discussed in the Supporting Information in Beretta et al.\cite{Beretta.Caironi.Sustainable}.}
    \label{fig:device_model}
\end{figure*}

The software is based on the non-linear model developed by Beretta et al.,\cite{Beretta.Caironi.Sustainable} solving the multi-physics problem illustrated in Fig.~\ref{fig:device_model} and described by the following system of non-linear equations: 
\begin{align}
    \label{eq:system1}
    \dot{q}_h &= h_{r,h}(T_{r,h}-T_{s,h}) \\
    \label{eq:system2}
    \dot{q}_h &= h_{s,h}(T_{s,h}-T_{h}) \\
    \label{eq:system3}
    \dot{q}_h &= h_{pni} (T_h-T_c) + \dot{q}_{\pi,h} - \dot{q}_j/2  \\
    \label{eq:system4}
    \dot{q}_c &= h_{pni} (T_h-T_c) + \dot{q}_{\pi,c} + \dot{q}_j/2 \\
    \label{eq:system5}
    \dot{q}_c &= h_{s,c}(T_{c}-T_{s,c}) \\
    \label{eq:system6}
    \dot{q}_c &= h_{r,c}(T_{s,c}-T_{r,c}) 
\end{align}
Here, $\dot{q}_{\pi,h}$, $\dot{q}_{\pi,c}$, and $\dot{q}_j$ are the Peltier and Joule heat per unit of device area, given by:
\begin{align}
\label{eq:system7}
    \dot{q}_{\pi,h} &= I_{cc}N\pi_{pn}\left(T_h\right)/A \\
    \dot{q}_{\pi,c} &=I_{cc}N\pi_{pn}\left(T_c\right)/A \\
    \dot{q_j} &=RI_{cc}^2/A
\end{align}
where $\pi_{pn}\left(T_h\right)$ and $\pi_{pn}\left(T_c\right)$ are the relative Peltier coefficients of the thermocouple at the hot- and cold-side temperatures, respectively, and $I_{cc}$ is the closed-circuit thermoelectric current. In a typical $\mu$TEG, $h_{s,h}$, $h_{pni}$ ($h_\mathrm{TEG}$ in the figure), and $h_{s,c}$ are determined solely by conduction, while $h_{r,h}$ and $h_{r,c}$ depend on the mechanism of thermal coupling and may include contributions from conduction, convection and radiation. The calculation of these coefficients is detailed in the Supporting Information of Beretta et al.\cite{Beretta.Caironi.Sustainable} Once the system of Eqs.~\ref{eq:system1}-\ref{eq:system6} is solved and the temperature profile of the device is determined, the software calculates the electric power $p$, the efficiency of conversion $\eta$, the electrical resistance $r$, the open circuit voltage $v_{oc}$, the short circuit current $i_{sc}$ and the closed circuit current $i_{cc}$ per unit of device area, according to the equations in Appendix \ref{sec:appendix}. 
The model is formulated as a parametric optimization problem. The software employs the finite difference method for numerical differentiation and uses SciPy’s \texttt{fsolve} to solve the system of equations, i.e. to determine the temperature profile. The initial conditions provided by the user include the temperatures $T_{r,h}$, $T_{s,h}$, $T_{h}$, $T_{c}$, $T_{s,c}$, and $T_{r,c}$, and the heat fluxes $\dot{q}_h$ and $\dot{q}_c$. Since \texttt{fsolve} is a local solver, the accuracy of the solution depends on the initial guess. Poor initial conditions may result in convergence to local minima or failure to find a solution. To improve convergence, it is good practice to choose the initial conditions on the temperatures such that $T_{r,h} >= T_{s,h} >= T_h >= T_c >= T_{s,c} >= T_{s,c}$, and to set the initial conditions on the heat flux such that $\dot{q}_h = h_{r,h}(T_{r,h} - T_{s,h})$ and $\dot{q}_c = h_{r,c}(T_{s,c} - T_{r,c})$. Following these conventions improves solver robustness and ensures the results are physically meaningful.

\section{User Interface}
The software's main window consists of three primary frames: (i) the input frame, (ii) the simulation frame, and (iii) the status bar frame.
\subsection{Input Frame}
The Input Frame allows users to define the model’s parameters, which are grouped into three categories across two tabs. Each parameter is assigned a specific data type and is restricted to a predefined range. If a user enters an invalid value, the background of the corresponding entry field turns red, preventing the simulation from being executed. Please refer to \cite{beretta_2024_12802052} for an overview of the parameters available in the current release, including their descriptions, units, data types, default values, and valid ranges. The model’s parameters can be reset to their default values, saved to, or loaded from a .json file. Parameters are only loaded if the file and its contents pass a validation check for data type and range. A template file is available on the project’s GitHub page for users to edit as needed.\cite{beretta_2024_12802052} All values are saved in SI units.

\subsection{Simulation Frame}
The Simulation Frame is designed to run simulations and display results. It includes various buttons and toggles that allow users to manage the simulation process and customize the visualization of results. These controls enable actions such as starting or clearing simulations, saving results to a file, adjusting the display of specific metrics, and switching between linear and logarithmic scales for the axes. For a complete up-to-date list of all the options available to the user, please refer to \cite{beretta_2024_12802052}.
	
\subsection{Status Bar Frame}
The Status Bar Frame provides real-time updates and status information, including errors and warnings encountered during the simulation process, as well as messages regarding invalid inputs and I/O operations.

\section{Representative Studies}
This section presents the results of two simulations for two $\mu$TEG configurations, both utilizing thermoelectric materials with the following properties: $\alpha_p = \SI{100}{\micro\volt\per\kelvin}$, $\alpha_n = \SI{-100}{\micro\volt\per\kelvin}$, $\sigma_p = \sigma_n = \SI{1e3}{\siemens\per\centi\meter}$, and $\kappa_p = \kappa_n = \SI{1}{\watt\per\meter\per\kelvin}$, yielding $zT = \alpha^2 \sigma / \kappa \approx 0.3$. The two configurations differ in their thermal coupling to the cold reservoir:
\begin{enumerate}
    \item Simulation 1: The $\mu$TEG is thermally coupled to the hot reservoir via direct mechanical contact, characterized by $h_{r,h} = \SI{1e6}{\watt\per\meter\squared\per\kelvin}$, and to the cold reservoir via heat exchangers, with $h_{r,c} = \SI{500}{\watt\per\meter\squared\per\kelvin}$. This configuration represents a standard $\mu$TEG design for low-power electronic applications, similar to those demonstrated by Böttner et al.\cite{Bottner.Bottner.2005}. 
    
    \begin{figure*}[t]
    \centering
    \includegraphics[width=1\linewidth]{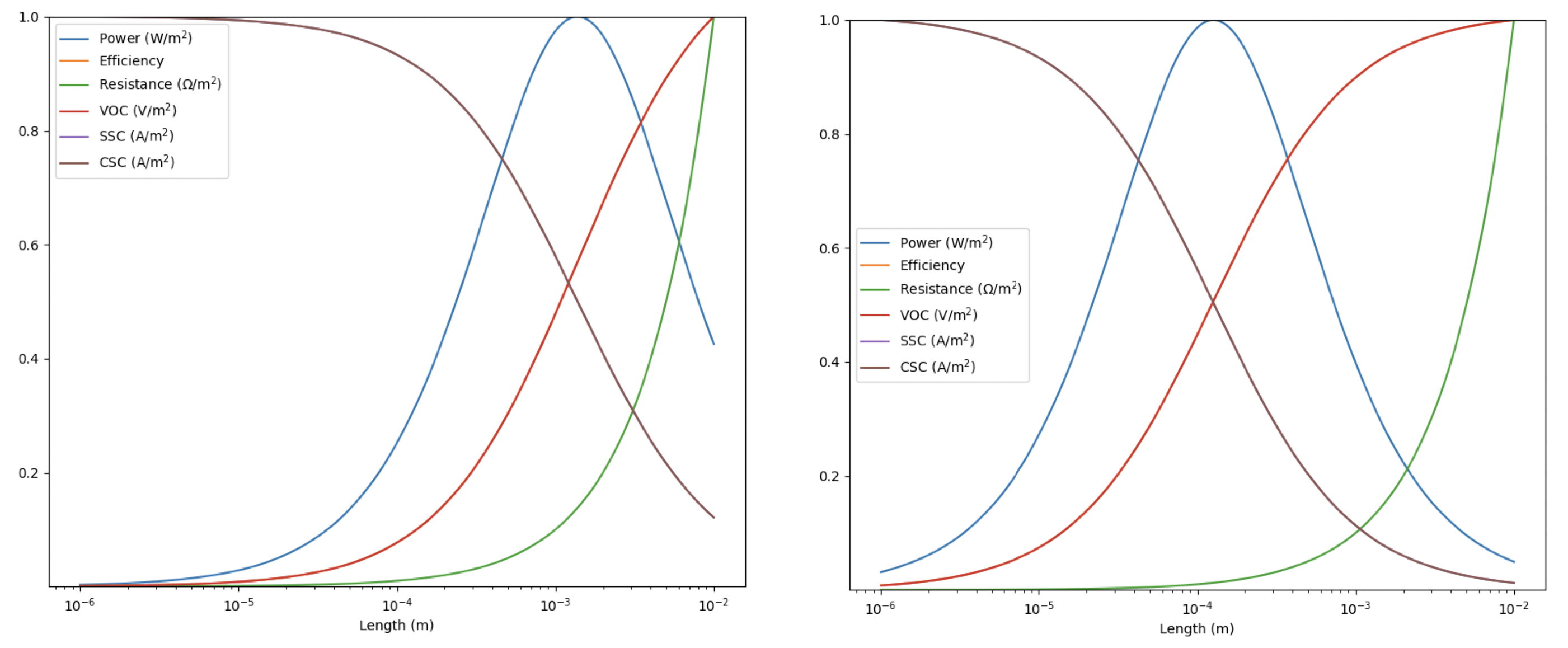}
    \caption{Left: normalized simulations results for a thermoelectric device thermally coupled to the hot reservoir via mechanical contact and to the cold reservoir via heat exchangers (Simulation n.1). Right: normalized simulation result for a thermoelectric device thermally coupled to both the hot and cold reservoirs by mechanical contact (Simulation n.2). Note that since the results are normalized, the short- and the closed-circuit currents overlap and therefore are indistinguishable.} 
    \label{fig:sims}
\end{figure*}
    \item Simulation 2: The $\mu$TEG is thermally coupled to both reservoirs via direct mechanical contact, with $h_{r,h} = h_{r,c} = \SI{1e6}{\watt\per\meter\squared\per\kelvin}$. While less common, this design shows significant potential for energy harvesting in $\mu$TEGs integrated into semiconductor devices, as it requires shorter thermocouple lengths for thermal matching. This makes it especially suitable for compact and efficient implementations.
\end{enumerate}
In both simulations, $\kappa_i = \SI{0.1}{\watt\per\meter\per\kelvin}$, while the thermal transfer coefficients of the substrates are set to $h_{s,h} = h_{s,c} = \SI{1e4}{\watt\per\meter\squared\per\kelvin}$, representing typical values for \SI{}{\milli\meter}-thick thermally conducting ceramic substrates. Both devices consist of thermocouples with a cross-sectional area of $\SI{100}{\micro\meter} \times \SI{100}{\micro\meter}$ and a fill fraction of \SI{50}{\percent}. Simulations were conducted for thermocouple lengths ranging from \SI{1}{\micro\meter} to \SI{1}{\centi\meter}, with $T_h = \SI{305}{\kelvin}$ and $T_c = \SI{300}{\kelvin}$ under condition of load-matching  ($m = 1$).
Fig.~\ref{fig:sims} shows the simulations results. The maximum power is achieved at thermocouple lengths that depend on the heat transfer coefficients. These lengths range from approximately \SI{1}{\milli\meter} for poor coupling with heat exchangers (Simulation n.1) to about \SI{100}{\micro\meter} for optimal direct mechanical coupling (Simulation n.2). As expected, the efficiency, resistance, and open-circuit voltage per unit of device area increase monotonically with thermocouple length, while the short- and the closed-circuit current per unit area decrease.
Interestingly, if the substrate heat transfer coefficients were increased by one or two orders of magnitude, the optimal thermocouple length would shift to the nanoscale range, making nanoscale thermoelectric generators (nTEGs), which exhibit promising thermoelectric quantum properties \cite{Dresselhaus.Gogna.Advanced.Materials.2007}, practically viable.

\section{Conclusions}
This work presents an open-source Python package designed to provide a user-friendly tool for simulating micro thermoelectric generators. Built on the lumped element model by Beretta et al.\cite{Beretta.Caironi.Sustainable}, the package features an intuitive graphical user interface, enabling users of all experience levels to perform basic simulations. The package is available via pip, executable from the command line, and its full source code is hosted on GitHub, ensuring transparency and allowing for further customization \cite{beretta_2024_12802052}. By simplifying the process of modeling thermoelectric systems, this Python package enables quick, efficient simulations, providing a foundation for more complex analyses and supporting further development in thermoelectric energy conversion.
\appendix
\section{Useful Equations}
\label{sec:appendix}

\subsection{The Open Circuit Voltage}
The thermoelectric voltage developed by one thermocouple is:
\begin{equation} \label{eq:v_pn}
    V_{pn} = \alpha_{pn} \left(T_h-T_c\right)
\end{equation}
The device open circuit voltage is given by the thermoelectric voltage developed by one thermocouple times the number of thermocouples $N$ in the device:
\begin{equation} \label{eq:V_oc}
    V_{oc} = N V_{pn} 
\end{equation}
The open circuit voltage per unit of device area is the open circuit voltage divided by the device area $A$. Given $N=nA$: 
\begin{equation} \label{eq:v_oc}
    v_{oc} = \frac{V_{oc}}{A}= nV_{pn}
\end{equation}
Note that since $N=A/A_{pni}$ and $N=nA$, then $n=1/A_{pni}$.

\subsection{The Electrical Resistance}
In Beretta et al. \cite{Beretta.Caironi.Sustainable}, the length $l$ of the \textit{p}- and \textit{n}-type thermoelectric legs is assumed to be the same, so that the electrical resistance of a thermocouple can be defined as:
\begin{align} \label{eq:r_pn}
    R_{pn} = \frac{\rho_p}{A_p}l + \frac{\rho_n}{A_n}l = \rho_{pn} \frac{l}{A_{pn}},
\end{align}
where $\rho_{pn} = (\rho_p / A_p + \rho_n / A_n)A_{pn}$ is the weighted average resistivity of the thermocouple. This formulation assumes an equivalent conductor with effective area $A_p+A_n$ and length $l$. Alternatively, one might consider a different effective conductor with an average cross-sectional area of $(A_p+A_n)/2$ and a length of $2l$ (i.e. twice the thermocouple length). In this case, the effective resistivity would be determined by:
\begin{equation}
    R_{pn} =
    \frac{\rho_p}{A_p}l + \frac{\rho_n}{A_n}l = \rho_{pn}' \frac{2 l}{\frac{A_{p}+A_n}{2}}
\end{equation}
which yields $\rho_{pn}'= (\rho_p / A_p + \rho_n / A_n)A_{pn}/4$, such that $4 \rho_{pn}'= \rho_{pn}$. Only this alternative formulation ensures that  $\rho_{pn}' \rightarrow \rho_p$ when $\rho_n = \rho_p$ and $A_p = A_n$. Regardless of the choice for the effective conductor, the underlying physics of the model remains unchanged. For the sake of consistency with the original work by Beretta et al.\cite{Beretta.Caironi.Sustainable}, this work makes use of the original definition. 

The device electrical resistance, or internal electrical resistance, is the electrical resistance of one thermocouple $R_{pn}$ times the number of thermocouples in the generator $N$:
\begin{equation} \label{eq:R_teg}
    R = N R_{pn} 
\end{equation}
The internal electrical resistance per unit of device area $r$ is the device electrical resistance divided by the device area $A$:
\begin{equation}
    r=\frac{R}{A}=nR_{pn}
\end{equation}

\subsection{The Thermoelectric Current}
For a single thermocouple, the short-circuit current $I_{pn}$ is given by Ohm's law:
\begin{equation} \label{eq:i_pn_sc}
    I_{pn} = \frac{V_{pn}}{R_{pn}}=\frac{\alpha_{pn}\left(T_h-T_c\right)}{\rho_{pn}l/A_{pn}}
\end{equation}
The device short-circuit current $I_{sc}$ is the ratio of the open-circuit voltage $V_{oc}$ to the device internal electrical resistance $R$. Since both $V_{oc}$ and $R$ scale linearly with the number of thermocouples $N$, the total short-circuit current is independent of $N$ and is equal to that of a single thermocouple:
\begin{equation} \label{eq:I_sc}
    I_{sc} = \frac{V_{oc}}{R} = \frac{N V_{pn}}{N R_{pn}} = I_{pn}
\end{equation}
The short-circuit current per unit of device area $i_{sc}$ is defined as the ratio of the short-circuit current to the device area. In formula:
\begin{equation} \label{eq:i_sc_density}
    i_{sc} = \frac{I_{sc}}{A} =\frac{f\alpha_{pn}\left(T_h-T_c\right)}{N\rho_{pn}l}
\end{equation}
Since the short-circuit current does not scale with the number of thermocouples in the device, for a given thermocouple design, the short-circuit current per unit of device is inversely proportional to the number of thermocouples.
In a more general case, the device is connected to an external load with resistance $R_{load}$, and the device closed-circuit current $I_{cc}$ is given by:
\begin{equation}\label{eq:I_cc}
    I_{cc} = \frac{V_{oc}}{R + R_{load}} = I_{sc}\frac{1}{1+m}
\end{equation}
where $m=R_{load}/NR_{pn}$. The closed-circuit current per unit of device area, $i_{cc}$, is again found by dividing the device closed-circuit current by the device area $A$:
\begin{equation}
    i_{cc} = \frac{I_{cc}}{A} =i_{sc}\frac{1}{1+m}
\end{equation}
This equation shows that the closed-circuit current per unit of device area depends on the fill fraction, on the load matching condition $m$, and on the number of thermocouples. 

\subsection{The Power}
The electrical power generated by the device and delivered to the load is given by:
\begin{equation} \label{eq:P}
    P = I_{cc}^2 R_{load}
\end{equation}
Using Eq.\ref{eq:I_cc} and Eq.\ref{eq:I_sc}, the power can be written as a function of $m$:
\begin{equation}\label{eq:P_expanded}
    P =  N\left( \frac{V_{pn}^2}{R_{pn}} \right) \frac{m}{(1+m)^2}
\end{equation}
The power per unit of device area, $p$, is defined as the power divided by the device area. Using Eq.\ref{eq:v_pn} and Eq.\ref{eq:r_pn}
\begin{equation} \label{eq:p}
    p = \frac{P}{A}=\frac{f\alpha_{pn}^2\left(T_h-T_c\right)^2}{\rho_{pn}l}\cdot\frac{m}{\left(1+m\right)^2}
\end{equation}
This result demonstrates that the power per unit of device area depends linearly on the fill fraction and on the load matching $m$.

\subsection{The Heat Transfer Coefficients}
The heat transfer coefficients of the substrates and the thermocouples depend solely on thermal conduction and are given by:
\begin{align}
    h_{s,h} &= \frac{\kappa_{s,h}} {l_{s,h}} \\
    h_{pni} &= \frac{\kappa_{pni}} {l} \\
    h_{s,c} &= \frac{\kappa_{s,c}} {l_{s,c}}
\end{align}
where $k_{pni}=\left(\kappa_p A_p + \kappa_n A_n + \kappa_i A_i\right)/A_{pni}$ is the weighted average thermal conductivity of a thermocouple unit. On the other hand, the heat transfer coefficients of the reservoirs depend on the mechanism of thermal coupling and may include contributions from conduction, convection and radiation.

\subsection{The Heat Fluxes}
The Peltier and Joule heat fluxes are:
\begin{align}
    \dot{Q}_{\pi,h}= & I_{cc}N\pi_{pn}\left(T_h\right)\\
    \dot{Q}_{\pi,c}= & I_{cc}N\pi_{pn}\left(T_c\right)\\
    \dot{Q}_{j}= & I_{cc}^2NR_{pn}
\end{align}
Using Eq.\ref{eq:I_cc}, Eq.\ref{eq:r_pn} and the Kelvin's relation $\pi_{pn}(T_h)=\alpha_{pn} T_h$ and $\pi_{pn}(T_c)=\alpha_{pn} T_c$, the heat fluxes per unit of device area can be written as:
\begin{align}
    \dot{q}_{\pi,h} &=\frac{\dot{Q}_{\pi,h}}{A}=\frac{f\alpha_{pn}^2(T_h-T_c)T_h}{\rho_{pn}l}\cdot\frac{1}{1+m} \\
    \dot{q}_{\pi,c} &=\frac{\dot{Q}_{\pi,c}}{A}= \frac{f\alpha_{pn}^2(T_h-T_c)T_c}{\rho_{pn}l}\cdot\frac{1}{1+m} \\
    \dot{q_j} &=\frac{\dot{Q}_j}{A} =\frac{f\alpha_{pn}^2(T_h-T_c)^2}{\rho_{pn}l}\cdot\frac{1}{\left(1+m\right)^2}
\end{align}
\subsection{The Efficiency}
The efficiency of energy conversion $\eta$ is the ratio between the power delivered to the load resistance and the heat flux absorbed from the heat reservoir. In formula: $$\eta=\frac{p}{\dot{q_h}}$$

\subsection{Optimal Conditions}
The maximum power output of a $\mu$TEG is achieved under the conditions of load matching and thermal matching. The load matching condition involves maximizing Eq.~\ref{eq:p} with respect to $m$. It is straightforward to show that the power is maximized when $m = 1$, meaning the load resistance should match the device resistance. 
On the other hand, the thermal matching condition is achieved by maximizing $\left(T_h - T_c\right)$. For the non-linear system of Eqs.~\ref{eq:system1}–\ref{eq:system6}, an analytical condition for thermal matching cannot be derived. However, Beretta et al.~\cite{Beretta.Caironi.Sustainable} demonstrated that, for the linearized model, the thermal matching condition corresponds to selecting a thermocouple length of $l = k_{pni}/h_{eq}$, where $h_{eq}^{-1} = h_{r,h}^{-1} + h_{s,h}^{-1} + h_{s,c}^{-1} + h_{r,c}^{-1}$ is the equivalent heat transfer coefficient. While this is only approximately correct, it provides a useful guideline for maximizing the temperature difference across the hot and cold ends of the thermocouples.

\setcounter{section}{0} 
\bibliographystyle{unsrturl}
\bibliography{main}  
\end{multicols}
\end{document}